\documentclass[lettersize,journal]{IEEEtran}

\ifCLASSOPTIONcompsoc
  \usepackage[nocompress]{cite}
\else
  \usepackage{cite}
\fi

\usepackage{booktabs}
\usepackage{comment}
\usepackage{mathtools}
\usepackage{multirow}

\usepackage{color,soul}
\usepackage{graphicx}
\usepackage{tikz}
\usepackage{amsmath}
\ifCLASSINFOpdf

\else

\fi

\newcommand\independent{\protect\mathpalette{\protect\independenT}{\perp}}
\def\independenT#1#2{\mathrel{\rlap{$#1#2$}\mkern2mu{#1#2}}}

\usepackage{url}

\usepackage{amsthm}
\theoremstyle{definition}
\newtheorem{definition}{Definition}[section]
\begin{document}
\title{Behind the Screens: Uncovering Bias in AI-Driven Video Interview Assessments Using Counterfactuals}
\author{Dena F. Mujtaba and Nihar R. Mahapatra \IEEEcompsocitemizethanks{Dena F. Mujtaba (email: mujtabad@egr.msu.edu) and Nihar R. Mahapatra (email: nrm@egr.msu.edu) are with the Department of Electrical and Computer Engineering, Michigan State University, East Lansing, MI, 48824}}

\markboth{}
{}

\maketitle

\begin{abstract} 
AI-enhanced personality assessments are increasingly shaping hiring decisions, using affective computing to predict traits from the Big Five (OCEAN) model. However, integrating AI into these assessments raises ethical concerns, especially around bias amplification rooted in training data. These biases can lead to discriminatory outcomes based on protected attributes like gender, ethnicity, and age. To address this, we introduce a counterfactual-based framework to systematically evaluate and quantify bias in AI-driven personality assessments. Our approach employs generative adversarial networks (GANs) to generate counterfactual representations of job applicants by altering protected attributes, enabling fairness analysis without access to the underlying model. Unlike traditional bias assessments that focus on unimodal or static data, our method supports multimodal evaluation—spanning visual, audio, and textual features. This comprehensive approach is particularly important in high-stakes applications like hiring, where third-party vendors often provide AI systems as black boxes.  Applied to a state-of-the-art personality prediction model, our method reveals significant disparities across demographic groups. We also validate our framework using a protected attribute classifier to confirm the effectiveness of our counterfactual generation. This work provides a scalable tool for fairness auditing of commercial AI hiring platforms, especially in black-box settings where training data and model internals are inaccessible. Our results highlight the importance of counterfactual approaches in improving ethical transparency in affective computing.
\end{abstract}


\section{Introduction}\label{sec:introduction} 
\IEEEPARstart{A}{}\textit{ffective computing}, often referred to as \textit{emotional AI}, is concerned with the study and development of systems capable of recognizing, interpreting, processing, and simulating \textit{human affects}, which encompass a wide range of human emotions and personality traits. Its applications in sectors like education, healthcare, and hiring—combined with rising investment—underscore its growing societal and economic influence \cite{booth2021integrating, caruelle2022affective}. The market for affective computing is projected to reach USD 123.3 billion by 2026, highlighting its significant economic potential \cite{caruelle2022affective}.

In this rapidly evolving landscape, advanced AI models like GPT-3.5 and GPT-4 have proven effective in various affective computing tasks, including personality prediction \cite{amin2023wide,amin2023will}. The Big Five/OCEAN model, which categorizes personality into the five dimensions of openness, conscientiousness, extraversion, agreeableness, and neuroticism, has become integral in areas such as recruitment and marketing, where aligning individual traits with organizational objectives and consumer behavior analysis is crucial \cite{john1991big, barrick1991big, mehta2020recent, caruelle2022affective}. The field has evolved from traditional methods like written or video assessments to modern, multimodal techniques that integrate visual, auditory, and textual data, allowing for more comprehensive and nuanced analyses \cite{zhao2019affective}.

\subsection{Motivation and Problem Background}
The increasing integration of affective computing models in high-stakes domains, such as recruitment, has intensified concerns over potential biases in these systems, leading to unfair outcomes. A notable instance of such bias was observed with Amazon's AI-based resume screening tool in 2017. This tool was withdrawn after it was found to systematically assign lower scores to resumes from women, due to their underrepresentation in the training data \cite{mujtaba2019ethical}. This problem is not limited to recruitment; similar biases have been identified in various AI applications, including search engines \cite{feng_has_nodate}, job recommendation systems \cite{carpenter2015google}, and image classification algorithms \cite{bril_2020}. Often, the root of these biases lies in the model training processes, which tend to prioritize prediction accuracy over fairness, thereby overlooking potential biases arising from the composition of the training dataset, the features employed, or the labeling practices used.

Building on these concerns, biases in affective computing models, particularly in personality prediction, pose additional challenges. For instance, the dataset used in the CVPR 2017 ChaLearn First Impressions (CLFI) competition was found to contain racial and gender biases in its labels \cite{escalante2020modeling,kaya2017multi}, and the winning model in the competition was later found to perpetuate these biases \cite{yan2020mitigating}. Such biases can arise from several sources, including the underrepresentation of specific groups within the training data, inherent biases in the original personality labels, and the features selected for training the models \cite{mujtaba2019ethical}. Additionally, it has been observed that personality traits can exhibit general trends across different subpopulations \cite{weisberg2011gender}, leading to potential bias when these trends are inferred by AI models.

Compounding this issue, other personality models like Myers-Briggs, often used in applications such as job performance prediction, have faced criticism for lacking empirical validity \cite{pittenger1993utility}. Furthermore, the field of emotion detection from films and photographs, another significant area in affective computing, has been challenged for its validity since emotions vary across cultures and require social context \cite{barrett2019emotional}. This underscores the critical need for careful training and regulation of such models to prevent the amplification of societal injustices.

\begin{figure*}[t]
	\centering
	\includegraphics[width=\textwidth]{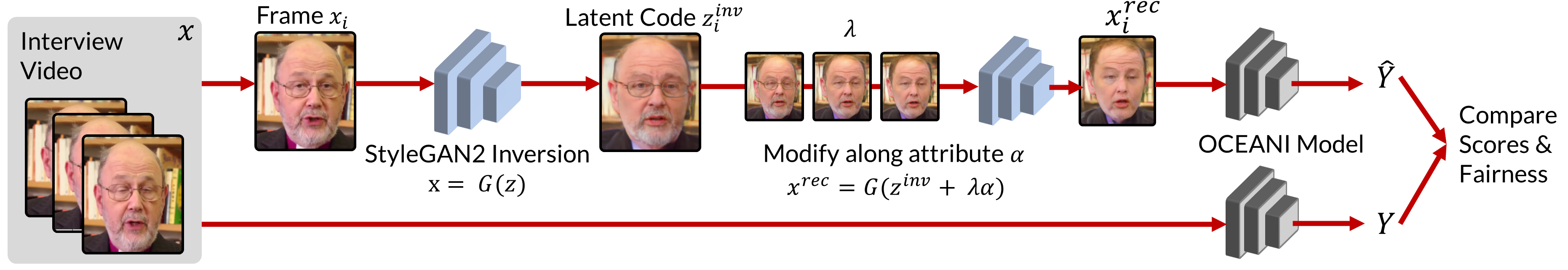}
	\caption{Overview of the proposed framework utilizing counterfactual videos for bias detection in affective computing models.}
	\label{fig:overview}
	\vspace*{-0.1in}
\end{figure*}

In response to these challenges, we present a novel bias measurement framework, depicted in Fig. \ref{fig:overview}, employing \textit{counterfactual videos} to detect and analyze biases in personality-driven models for job candidate recommendation. Central to our approach is the principle of \textit{counterfactual fairness} \cite{kusner2017counterfactual,pearl2016causal}. This concept posits that a model is deemed fair if its predictions for an individual in the real world would be consistent with those in a counterfactual world where the individual's protected attributes (such as gender or ethnicity) are different. Formally, this can be expressed as follows:

\begin{definition}[Counterfactual Fairness]\label{def-counterfactual}
\textit{A predictor $\hat{Y}$ is counterfactually fair if, given protected attributes $A$ (e.g., gender, age), general attributes $X$, and model prediction $\hat{Y}$, the probability of $\hat{Y}$ remains unchanged when $A$ is counterfactually altered. Formally, given a causal model $(U, V, F)$, where $V$ is the set of observable variables (including both protected attributes $A$ and non-protected attributes $X$), $U$ represents unobserved background factors influencing $V$, and $F$ is a set of functions $\{ f_1, \dots, f_n \}$ describing the causal relationships between these variables \cite{kusner2017counterfactual}, counterfactual fairness is defined as:
\begin{equation}
\begin{multlined}
    P(\hat{Y}_{A \leftarrow a} (U) = y \mid X = x, A = a) =  \\
    P(\hat{Y}_{A \leftarrow a'} (U) = y \mid X = x, A = a')
\end{multlined}
\end{equation}
for all possible values of $y$ and for all attainable counterfactual values $a'$ of $A$ \cite{kusner2017counterfactual,pearl2016causal}. In simpler terms, a model is counterfactually fair if changing an individual’s protected attribute(s) alone does not affect its prediction.}
\end{definition}

Distinct from traditional fairness metrics, counterfactual fairness evaluates model predictions post-training, a crucial consideration in contexts like hiring and personality prediction where end-users, such as companies and HR managers, often rely on external services and lack direct access to the underlying model or its training data \cite{raghavan2020mitigating}. For instance, consider a scenario where a job candidate receives a lower score in an automated interview. Counterfactual analysis allows us to determine whether factors like the candidate's age unjustly influenced this score, as opposed to their actual qualifications and abilities. This becomes particularly relevant when considering potential biases in ground-truth labels, as standard metrics like predictive rate parity or accuracy parity assume unbiased labels, making them less effective in identifying inherent biases in the data. Label biases are also evident in datasets like the CLFI, where initial personality assessments made by humans have shown bias against certain gender and racial groups \cite{junior2021person}. Therefore, a counterfactual approach is instrumental for conducting thorough bias audits in modern AI hiring systems and other similar applications—especially in situations where the AI system is externally provided and model internals or training data are not accessible to end users.

Our framework relies on \textit{counterfactual face attribute manipulation}, wherein a generative adversarial network (GAN) model is used to create alternative reality videos of individuals \cite{denton2019image}. We employ this method to generate counterfactual video interviews of job candidates by modifying their protected attributes (e.g., age or gender) to represent an opposite group. These new videos, along with the original unmodified videos, are then input to a personality and interview score predictor model. The observed changes in personality and overall interview score predictions are used to assess bias. A model is considered fair if it predicts similar scores for both the original and counterfactual videos. We also combine this approach with existing fairness metrics to study changes in regression applications (e.g., personality trait scores) and classification applications (e.g., job candidate selection based on personality traits).

\subsection{Key Contributions}

Our research advances personality prediction and other areas of affective computing by making several key contributions:

\begin{itemize}
    \item \textbf{Counterfactual Bias Measurement:} We introduce a new framework that utilizes counterfactual examples, leveraging GANs and latent space manipulation of protected attributes to investigate biases. While previous studies in fairness have employed GANs for generating images to enhance fairness during training and evaluation \cite{sattigeri2018fairness,ramaswamy2021fair}, or to analyze model score sensitivity \cite{denton2019image}, our approach is distinctive in its application to pre-trained models in affective computing. It performs isolated manipulation of protected attributes to create counterfactuals 
    without retraining or modifying the original model. Our method enables fairness auditing of black-box systems—where only input-output access is available—a common reality in commercial AI hiring tools.

    \item \textbf{Fairness in Video-Based and Multimodal Models:} We explore fairness in video-based models, which introduces challenges not present in static image analysis—such as handling multimodal inputs (visual, auditory, and textual), separating facial and scenic features, ensuring temporal consistency and realism in facial movement, and dealing with the computational complexity of video model training. Our research examines multimodal fairness using a state-of-the-art affective computing model for personality prediction and a personality-driven job candidate selection system based on the Big Five (OCEAN) model, which includes Openness (O), Conscientiousness (C), Extraversion (E), Agreeableness (A), and Neuroticism (N) traits \cite{barrick1991big,mehta2020recent}. This allows us to evaluate both regression (e.g., trait scores) and classification (e.g., candidate selection) model behaviors.

    \item \textbf{Validation via Attribute Classifier:} We validate our counterfactual video generation framework using a separately trained protected attribute classification model. This model confirms that counterfactual videos are perceived as belonging to the intended counterfactual group, supporting the effectiveness of our GAN-based attribute editing method. As both the attribute classifier and the personality dataset rely on first impressions, we clarify their scope and intended ethical use in Sec. \ref{S-ethics}.

    \item \textbf{Auditing Without Model Access:} Our framework enables fairness assessment in black-box systems using only input-output access—without requiring knowledge of model internals or training data.
    This is a common and growing scenario in commercial AI-based hiring platforms, which often offer video assessment tools as third-party services. Our method empowers users to perform fairness evaluations based solely on model inputs and outputs.
\end{itemize}

The remainder of this paper is organized as follows. In Section \ref{S-related}, we provide a brief literature review of fairness in AI and the use of GANs in AI fairness. 
In Section \ref{S-methods}, we detail the methodologies employed in our research, including the dataset used, the models we developed, our approach to generating counterfactual videos, and the fairness metrics applied. Section \ref{S-ethics} delves into the ethical considerations pertinent to our work and its intended applications. Section \ref{S-results} presents our experiments and the results obtained. Finally, Section \ref{S-conclusion} offers concluding remarks and discusses potential areas for future work.

\section{Background and Related Work}\label{S-related} 
Before measuring bias, the concept of ``fairness'' needs to be defined in the context of AI. The following sections define fairness as described in past literature on AI (Sec. \ref{S-related-fairness}) and outline previous methods that use GANs for this purpose and how they relate to our work (Sec. \ref{S-related-GANs}).

\subsection{Fairness in Artificial Intelligence for Hiring}\label{S-related-fairness}
Many researchers have proposed definitions of fairness based on past anti-discrimination legislation, such as the Civil Rights Act of 1964, which prohibits unfair treatment of individuals based on their protected attributes \cite{act1964civil}. Similarly, many frameworks for defining ethical AI, such as those created by the National Institute of Standards and Technology (NIST) \cite{nistproposalAI2021}, the Federal Trade Commission (FTC) \cite{ftc-2021}, and the U.S. Equal Employment Opportunity Commission (EEOC) \cite{us_eeoc}, outline the need for fair, transparent, and responsible AI to prevent adverse impacts on decisions made by these systems. Only recently, in 2023, AI-related legislation went into effect to protect the sharing of personal information through AI, regulate its use in hiring, and require transparency in decision-making \cite{katrina2023StateofStateAI}. Several definitions of fairness stemming from these standards have been established to measure biases in AI models, the simplest being fairness through unawareness:

\begin{definition}[Fairness Through Unawareness]\label{def-ftu}
\textit{Given a set of protected attributes $A$ of an individual and a set of other observable features $X$, any given model $C$ is considered fair as long as the protected attributes $A$ are not explicitly used in its decision-making process. This is explicitly defined by any mapping $\hat{Y}: X \rightarrow Y$ that excludes $A$, where $\hat{Y}$ is the model $C$ outcome prediction, and $Y$ is the outcome to be predicted \cite{kusner2017counterfactual}.}
\end{definition}

This concept has been used to define metrics that quantify fairness in both regression and classification applications. For example, in regression problems, fairness through unawareness is commonly modeled through the statistical independence of the protected attribute in model predictions \cite{barocas2017fairness}. Formally, this is defined as follows:

\begin{definition}[Independence \cite{barocas2017fairness}]
\textit{Independence states that the joint distribution of a sensitive attribute $A$ and classifier predictions $Y$ must have zero mutual information, meaning they are independent of one another. This is expressed as $Y \independent A$. To assess independence, one could use the estimated mutual information (MI) \cite{sklearnmi}, wherein MI $=0$ if and only if two random variables are independent, and larger values indicate a higher dependency.}
\end{definition}

In the case of classification, such as job candidate selection, a frequently referenced metric is \textit{disparate impact}, which measures the relative rate at which individuals from protected and unprotected groups receive positive outcomes. Previous legislation by the EEOC has relied upon this metric to define adverse impact, stating that the selection rate for a minority group should be at least 80\% of the selection rate for the majority group \cite{equal1979adoption, barocas2016big}. However, this threshold has recently been scrutinized due to significant differences in selection rates still persisting \cite{cohenmilstein_ai_bias}, as ideally, a fairer model would result in similar hiring outcomes across all groups. Formally, disparate impact is defined as follows:

\begin{definition}[Disparate Impact \cite{mujtaba2019ethical,barocas2017fairness}]
\textit{Disparate impact compares the proportion of individuals from the unprivileged group to the privileged group who receive a positive outcome, specifically given by:
\begin{equation}
    \frac{P\{C=1 \mid D=unprivileged\}}{P\{C=1 \mid D=privileged\}},
\end{equation}
where $C=1$ represents a positive hiring outcome predicted by the model, and $D$ denotes the group, either unprivileged or privileged. For a model to be considered fair, this measure should ideally be equal to 1. However, a threshold of 0.8 is commonly accepted in practice.}
\end{definition}

Biases identified by these fairness metrics can stem from several causes, including a skewed sample in the training dataset, bias in the data labels, or even the features selected for model training \cite{mujtaba2019ethical}. For example, academic credentials are commonly used in hiring, but if members of a job candidate's protected group graduate at a disproportionately higher or lower rate from certain colleges, a model may learn to incorrectly classify job candidates from that group, even if they are qualified for the job \cite{mujtaba2019ethical}. Therefore, it is essential to have a robust approach to measuring fairness, which we seek to address for affective computing models.

\subsection{Generative Adversarial Networks for Fairness}\label{S-related-GANs}

For a detailed review of AI bias mitigation and measurement methods, we point readers to the survey by Mehrabi et al. \cite{mehrabi2021survey}. Specifically with GANs, there is current work on bias mitigation and measurement related to our approach which we outline. A majority of these methods fall into the pre-processing and in-processing stages. First, for pre-processing, the approach by Sattigeri et al. \cite{sattigeri2018fairness}, FairGAN \cite{xu2018fairgan}, and others \cite{morales2020sensitivenets,georgopoulos2021mitigating} use GANs to modify a model's training dataset to meet fairness constraints. Next, the FairFaceGAN method uses image-to-image translation to fairly map images to a GAN's latent space for modification \cite{hwang2020fairfacegan}. Moreover, the approach by Park et al. \cite{park2021learning} learns fair representations of facial images to classify facial expressions. In comparison to these methods, in-processing approaches have used adversarial learning \cite{edwards2015censoring,yan2020mitigating}, and modification of a model's loss function to meet fairness constraints while training. Furthermore, the approach by Pena et al. \cite{pena2020bias,morales2020sensitivenets}, similar to ours which predicts a job interview score, creates a fair classification model using resumes and multi-modal data of job candidates profiles. 

Similarly, others have worked on counterfactual fairness with GANs. First, the work by Li et al. \cite{li2021discover} discovers the attribute from photos which contributes to biased model predictions; their approach uses a three-stage process, in which humans speculate potential biases, collected label testing images, and analyze a classifier's predictions. Similarly, the approach by Ramaswany et al. \cite{ramaswamy2021fair} uses GAN latent space manipulation for de-biasing. Likewise, the work by Luo et al. \cite{luo2023zero} and Prabhu et al. \cite{prabhu2023lance} follow a similar technique, though not necessary using just human images. Next, Dash et al. \cite{dash2022evaluating} use a structural causal model to generate counterfactuals. Last, the work by Denton et al. \cite{denton2019image} define counterfactual sensitivity analysis and analyze changes in model predictions after changing attributes. A similar work to this, by Balakrishnan et al. \cite{balakrishnan2021towards}, uses transects, or generated arrays of images, with a similar analysis approach. 

In comparison to past approaches, we focus on pre-processing job candidates' interview videos after a model has already been trained. Often, candidates and employers do not have access to the screening model and cannot assess the fairness of its decisions. Therefore, we propose an approach that generates a fairer version of the interview data. 
Unlike our approach, prior methods have largely focused on static images and have not addressed the unique challenges of video data, such as temporal consistency, audio integration, and multimodal representation.

\section{Materials and Methods}\label{S-methods}

Our work consists of three experiments to implement and evaluate our framework. First, in \textbf{Experiment 1}, we train a personality prediction model (described in Sec. \ref{S-methods-model}) that takes an input video of a hypothetical job candidate and outputs their OCEAN personality traits and an interview score, referred to as OCEANI. Results from this experiment compare the model's accuracy across different personality traits and protected groups.

Next, in \textbf{Experiment 2}, we apply our counterfactual facial attribute manipulation technique (Sec. \ref{S-Counterfactual}) to modify job candidate videos by changing a protected attribute (e.g., reducing their perceived age) and then input these new videos into the personality prediction model of Experiment 1. The newly obtained scores are then compared with the original scores to analyze bias and its downstream effects on job candidates from different protected groups. Regression metrics study fine-grained changes in OCEANI predictions both between different groups and within each individual. For this, we use the Independence metric (previously described in Sec. \ref{S-related-fairness}) to study bias in the changes of OCEANI scores. However, since many downstream applications such as resume selection and ranking are classification problems, we also rely on the predicted interview score and simulate a candidate selection model. This is done in two ways: first, by selecting the top $N$ candidates based on their interview score, and second, by selecting only candidates with an interview score above an $\epsilon$ threshold. Then, we use the metric of Disparate Impact (previously described in Sec. \ref{S-related-fairness}), which is based on hiring legislation, to quantify selection bias.

Lastly, in \textbf{Experiment 3}, we evaluate our method for counterfactual generation by training a separate model for predicting job candidates' protected attributes. We use this to assess the effectiveness of our counterfactuals, where a decline in the protected attribute model's accuracy after attribute manipulation indicates successful changes to the individuals' protected attributes. We also highlight our intent for the ethical use of this model and our overall framework in Sec. \ref{S-ethics}. Our model architecture, counterfactual facial attribute manipulation approach, and dataset used are described next.

\subsection{Dataset}\label{S-dataset}

We use the CLFI dataset \cite{escalante2020modeling,kaya2017multi}, one of the largest datasets available for personality prediction \cite{mehta2020recent}. It contains 10,000 videos of individuals speaking to a camera, each 15 seconds long, with labels for their OCEAN personality trait scores and a job interview score (i.e., the likelihood of being invited back for an interview). We refer to all these scores collectively as OCEANI, where I represents the interview score. These labels, obtained through Mechanical Turk and pairwise comparisons between videos, represent first impressions. Each individual's scores for each OCEANI dimension range from 0 to 1, with 1 being the highest score and 0 the lowest. The distribution of this dataset among the protected attribute groups is illustrated in Table \ref{tbl:dataset}. While the gender attribute is fairly balanced across the dataset, there is an imbalance in ethnicity and age, which may contribute to bias.

We note that for the age attribute, we remove videos without an age label when analyzing bias. 
For the remaining videos, we group individuals into two age categories based on legal protection under the Age Discrimination in Employment Act (ADEA) \cite{hiringact}: those aged 40 or older (protected group) and those under 40 (reference group).

\begin{table}[t]
  \centering
  \caption{Dataset Distribution Per Protected Attribute Group, Before and After Filtering. The first row (``Before'') contains videos from the original dataset. The second row (``After'') contains videos after filtering out those in which individuals were not facing the camera and could not undergo counterfactual facial manipulation. Demographic groups are abbreviated as follows: Female (F), Male (M), Caucasian (Cau.), African-American (Af.Am.), Asian (Asi.), individuals at or above the age of 40 ($\text{A} \ge 40$), and individuals below the age of 40 ($\text{A} < 40$).}
  \label{tbl:dataset}
  \renewcommand{\arraystretch}{1.2}
  \setlength{\tabcolsep}{4.5pt}
  \begin{tabular}{lccccccc}
    \toprule
    \textbf{Stage} & \multicolumn{2}{c}{\textbf{Gender}} & \multicolumn{3}{c}{\textbf{Ethnicity}} & \multicolumn{2}{c}{\textbf{Age}} \\
    \cmidrule(lr){2-3} \cmidrule(lr){4-6} \cmidrule(lr){7-8}
    & F & M  & Af.Am. & Cau. & Asi. & $\text{A} < 40$ & $\text{A} \ge 40$\\
    \midrule
    Before & 5,462 & 4,538 & 1,071 & 8,598 & 331 &  8,214 & 384\\
    After  & 3,916 & 2,764 & 842 & 5,572 & 266 &  5,346 & 226\\
    \bottomrule
  \end{tabular}
\end{table}

\begin{figure*}[t]
	\centering
	\includegraphics[width=\textwidth]{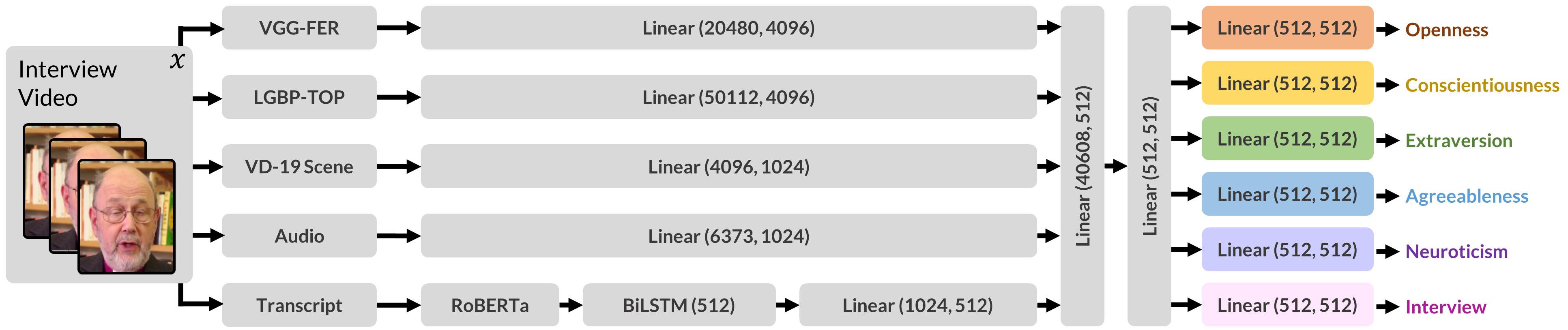}
	\caption{Architecture of the multi-task deep neural network model (MTDNN model) \cite{mujtaba2021multi} used for personality and interview score prediction. }
	\label{fig:model_architecture}
	\vspace*{-0.1in}
\end{figure*}

\subsection{Personality Prediction Models}\label{S-methods-model}
Our experiments utilize two models. The first is a multi-task deep neural network (\textbf{MTDNN}) model based on the state-of-the-art architecture described by Mujtaba and Mahapatra \cite{mujtaba2021multi}, and is used in Experiments 1 and 2. The second is a multi-task model for predicting protected attributes, which we refer to as the \textbf{P-Attribute} model, and is used in Experiment 3. The MTDNN model and its variants follow a training procedure and architecture similar to that described in \cite{mujtaba2021multi}. An overview of both models, including their architecture and training procedure, is provided next.

\subsubsection{OCEANI predictor model (MTDNN model)}
The architecture of our OCEANI predictor model, or MTDNN model, is depicted in Figure \ref{fig:model_architecture}.
This model is a multi-task neural network where all inputs are combined as input to shared hidden layers. Each OCEANI score then has its own task-specific layer to output final predictions. The advantage of this method is its ability to share information across the shared layers and optimize task-specific predictions at the task-specific layers. Moreover, this reduces training time, as only one model is needed for training, rather than separate models for each score and feature extraction. Additionally, information learned from video features such as the face, scene, audio, and text is used to infer all OCEAN traits and the interview score.

Overall, the features input to our model consist of face, scene, audio, and text features extracted from interview videos. These are modeled after the BU-NKU model feature extraction method, presented by Kaya et al. \cite{kaya2017multi}, for our approach. First, faces of individuals in the videos are aligned using IntraFace, and facial features are extracted using a pretrained VGG-Face network \cite{de2015intraface,kaya2017multi}. Second, scene features (i.e., the individual's surroundings in the video) are extracted using the VGG-VD-19 network, a model trained for object recognition that detects object-like parts in the first frame of the video \cite{kaya2017multi,yan2020mitigating}. Third, deep facial features are compared and combined for effective emotion recognition using a spatio-temporal descriptor, Local Gabor Binary Patterns from Three Orthogonal Planes (LGBP-TOP), which has been shown to be effective for models to predict emotions \cite{kaya2017multi}. Lastly, acoustic features (e.g., pitch and Mel Frequency Cepstral Coefficients) are extracted using the OpenSMILE toolkit \cite{eyben2010opensmile}. These features, in prior model results, have shown to provide high accuracy towards the prediction of OCEANI traits, indicating they are useful for our counterfactual analysis.

In addition to these features, we also integrate the audio transcript of each individual, as prior work has shown personality can be inferred from language use \cite{mehta2020recent}. The CLFI dataset provides high-quality human-transcribed video transcripts \cite{cvpr}. Each transcript is tokenized and embedded using RoBERTa, a state-of-the-art pretrained language model. The resulting embeddings are passed through a bidirectional long short-term memory (BiLSTM) layer and fused with the other modality features in the shared layers of the MTDNN model for the prediction of final OCEANI scores.\footnote{RoBERTa improves upon BERT by optimizing its pretraining setup, including batch size, training duration, sequence length, and masking strategy \cite{devlin2018bert, liu2019roberta}.}

All variations of our MTDNN model were implemented in PyTorch \cite{paszke2019pytorch} and trained on one NVIDIA Tesla-P100 GPU. We use a batch size of $N=10$ and the stochastic gradient descent optimizer with a learning rate of $0.001$ and momentum $0.9$. The learning rate is decayed by a factor of $\gamma=0.1$ every 5 epochs. In each hidden layer, weights are initialized using Xavier initialization, and batch normalization is applied \cite{paszke2019pytorch}. Next, a dropout rate of $p=0.2$ is used. Each hidden layer uses the rectified linear unit (ReLU) activation function, and the output of each task-specific layer uses a Sigmoid activation function. The loss function used for each task in our MTDNN model is the mean squared error (MSE) criterion (squared L2 norm) as described in \cite{mujtaba2021multi}. Then, to evaluate our model, we use the mean absolute error criterion, given by:
\begin{equation}
\frac{1}{N}\sum^N_{n=1}|x_n - y_n|,
\end{equation} for an input $x$, label $y$, and batch size $N$ \cite{paszke2019pytorch}. Each model is trained for 15 epochs, and we use early stopping to select model weights with the highest validation accuracy. We use five-fold cross validation in each experiment.

\subsubsection{Protected attribute predictor model (P-Attribute model)}\label{S-attribute-model}
The overall architecture of this model is identical to that of our MTDNN model, with the exception of its initial input layers. Instead of multiple input layers for different features, we input a sequence of $64 \times 64$ pixel image frames, into a RegNet \cite{radosavovic2020designing} model and long short-term memory (LSTM) layers, specifically 3 stacked LSTM layers with a hidden size of 128. RegNet is a model consisting of convolutional neural network (CNN) layers based on a design space of simple and consistent network blocks. The concept behind RegNet is to create a network design space consisting of simpler models that perform as well as state-of-the-art image classifiers (e.g., EfficientNet), while reducing computational costs. Each block in RegNet builds upon ResNet (Residual Networks), a model that introduced deep residual learning, allowing shortcut connections between network layers \cite{he2015deep}. We use the RegNetX-16GF configuration, where 16GF indicates $16 \times 10^9$ FLOPS used for training. With RegNet, we utilize transfer learning, enabling knowledge from a pretrained model, such as object detection learned on the ImageNet dataset, to be transferred to our regression problem, thereby improving performance. 

These input layers are merged and input to task-specific layers for predicting ethnicity, gender, and age. These predictors are linear layers with input sizes similar to the OCEANI task-specific layers; their output sizes match the number of classes for each attribute. A sigmoid activation function is not used for the output of each task-specific layer since our model performs classification. This model was also implemented in PyTorch, following a similar training procedure to our MTDNN model, with modifications for multi-class classification. We also use five-fold cross-validation for training. However, since the number of individuals per minority group is lower than that of the majority group, we modify our loss function to the following unreduced cross-entropy loss $l(x,y)=L=\{l_1, \dots ,l_N\}^T$, where the loss for every $n$th item in the batch of size $N$ is:
\begin{equation}
    l = - \sum^C_{c=1}w_c \log \frac{e^{x_{n,c}}}{\sum^C_{i=1}e^{x_{n,i}}}y_{n,c},
\end{equation}
where $x$ is the input image, $y$ is the target label, $C$ is the total number of classes, and $w$ is a weight vector for individual weights applied per item in the batch. For both the ethnicity and age categories, we apply weights according to the ratio of the number of individuals in the majority group to the number of individuals in the minority group. For gender, both groups were weighted equally since there was a nearly equal number of records for both groups.

\subsection{Counterfactual Video Generation}\label{S-Counterfactual}
Our counterfactual video generation process contains four stages. First, we extract each video's frames and align the faces of each individual to the center of the image using Python's dlib library \cite{king2009dlib} and the FFHQ pre-trained face extraction model \cite{karras2019style}. From each frame, we produce $256 \times 256$ and $64 \times 64$ pixel images, with the $64 \times 64$ images being used for feature extraction and model training, and the $256 \times 256$ images used for modification, which are later resized to $64 \times 64$ images. For our MTDNN model, we extract facial, movement, and scenic features from these images using the procedure described by Kaya et al. \cite{kaya2017multi}. Next, to synthesize modified image frames, we invert each frame image back to our GAN's latent space (i.e., the GAN's search space of latent codes generated from a fixed prior distribution used during training). The GAN model and inversion method we use are further described later. Afterwards, the inverted images, which map to our GAN model's latent space, can be modified along a learned semantic boundary (e.g., gender or ethnicity). These modifications target a single set of attributes that map along a direction in the latent space, which is advantageous because unrelated attributes remain unchanged during image modification. Finally, the modified image frames are used as counterfactuals for all our experiments.

\subsubsection{Generative adversarial networks}\label{S-GAN}
Generative adversarial networks (GANs) are a powerful framework for \textit{generative modeling} and have been used to create models to generate images, sound, and text \cite{goodfellow2016nips,goodfellow2014generative}. In GANs, two models compete during training: a \textit{generator} and a \textit{discriminator}. The generator, $G$, creates data samples $x$ that aim to be indistinguishable from real data based on the training data distribution $p_g$ \cite{goodfellow2014generative}. The generator competes against an adversary, the discriminator, $D$, that classifies these samples as either real (i.e., part of the original dataset) or fake (i.e., generated by $G$) \cite{goodfellow2016nips}.

An input noise $p_z(z)$ from the data space $G(z; \theta_g)$, where $\theta_g$ are model parameters, is fed into the generator. The discriminator, $D$, is trained to maximize the accuracy of $D(x; \theta_d)$, the probability that the data sample $x$ is real and not generated by $G$. Simultaneously, $G$ is trained to minimize $\log(1-D(G(z)))$, reducing the discriminator's ability to correctly classify the generated data. This results in the following optimization problem \cite{goodfellow2014generative}:
\begin{equation}
\begin{split}
    \min_G \max_D V(D,G) =\ & \mathbf{E}_{x\sim p_{data}(x)} [\log D(x)] \ + \\ &\mathbf{E}_{z\sim p_{z}(z)} [\log(1-D(G(z)))],
\end{split}
\end{equation} 
where $p_{data}$ is the distribution of the training data.

GANs have many advantages over standard generative models, including the ability to generate highly realistic images. For our approach, we use StyleGAN to modify video interviews.

\subsubsection{StyleGAN}
StyleGAN is a GAN architecture introduced in 2019 that re-designs the generator model to include mapping and style networks, resulting in the generation of photo-realistic and high-quality images \cite{karras2019style}. It has been used for image-to-image translation \cite{viazovetskyi2020stylegan2}, face generation \cite{shi2021lifting}, and in medical applications \cite{su2020pre}. In contrast to traditional GAN models, where only a single point from the latent space is given as input, StyleGAN adds two additional sources of randomness: a mapping neural network and noise layers. First, a latent code $\mathbf{z}$ is given from the latent space $\mathbf{z} \in \mathcal{Z}$ \cite{karras2019style, karras2020analyzing}. This code is then input to StyleGAN's mapping neural network $f$, an 8-layer MLP, which outputs $\mathbf{w} \in \mathcal{W}$, an intermediate latent space \cite{karras2019style, karras2020analyzing}. The vector $\mathbf{w}$ is input to the StyleGAN generator model, where spatially invariant styles $y = (y_s, y_b)$ can be computed for each image, with $y_s$ used for scaling and $y_b$ for style transformation as bias \cite{karras2019style, karras2020analyzing}. The generator model in StyleGAN provides control over fine details and has the capability to create more realistic images due to the intermediate latent space and its unique architecture.

Between each convolution layer in the generator network, a control adaptive instance normalization (AdaIN) operation is applied, as described by \cite{karras2019style}:
\begin{equation}
\text{AdaIN}(x_i, y) = y_{s,i} \frac{x_i - \mu(x_i)}{\sigma(x_i)} + y_{b,i},
\end{equation}
where $x_i$ is a feature map normalized separately and scaled with the style from $y$, supporting style transfer.

In 2020, StyleGAN was extended to StyleGAN2, significantly improving the quality of generated images by replacing normalization in the generator model with demodulation, thereby removing characteristic artifacts \cite{karras2020analyzing,Karras2020ada}. For our work, we use the StyleGAN2 implementation in PyTorch \cite{Karras2020ada}, pre-trained on the Flickr-Faces-HQ dataset, which contains 1024x1024 resolution facial images.\footnote{\url{https://github.com/NVlabs/ffhq-dataset}} Although StyleGAN2 was further extended to StyleGAN3 in 2021, we use StyleGAN2 for all experiments due to its superior capability for style mixing. Style mixing was disabled in StyleGAN3 training  due to the focus of the model, which resulted in poorer performance for our purposes compared to StyleGAN2 \cite{karras2021alias}.

\subsubsection{Attribute manipulation}
To modify each individual frame of a video, we first need to map the image to StyleGAN's latent space, a process known as \textit{GAN inversion}. This is challenging due to the size and complexity of a GAN's latent space, where standard optimization methods (e.g., gradient descent) for searching a latent code are too time-consuming and impractical. Therefore, existing methods train an encoder to map images to latent codes and a decoder to reconstruct the original image \cite{zhu2020domain}. However, past work has mainly focused on reconstruction using solely image pixels, leaving out semantic knowledge encoded in the latent space that would assist in more accurate image editing.

To address this, we use the In-Domain GAN Inversion approach by Zhu et al. \cite{zhu2020domain}. This method offers two advantages: (i) the use of a domain-guided encoder to invert images, ensuring all codes are in-domain (i.e., they also encode semantic knowledge) for better editing, and (ii) the use of instance-based domain-regularized optimization for image reconstruction, which supports starting at an ideal point to avoid local minima and create more realistic images \cite{zhu2020domain}. Thus, for each image $\mathbf{x}$, instead of the standard generation method where $\mathbf{x} = G(\mathbf{z})$ and $\mathbf{z}$ is the corresponding latent code for the image, Zhu et al.'s inversion method inputs $\mathbf{x}$ to an encoder $E$, resulting in $\mathbf{z}^{inv}$ for modification \cite{zhu2020domain}.

For image modification, we use the approach by Shen et al. \cite{shen2020interpreting}, Interpreting Face GAN (InterFaceGAN). In this approach, attribute boundary vectors $\alpha$ are inferred from a GAN's latent codes and corresponding image attribute annotations. Therefore, for any obtained latent code $\mathbf{z}^{inv}$, a new image is reconstructed by $\mathbf{x}^{rec} = G(\mathbf{z}^{inv} + \lambda\alpha)$, wherein $\alpha$ is the learned boundary and $\lambda \in [-\infty, \infty]$ is the direction of attribute modification (as $\lambda$ increases, the attribute prominence increases, and as $\lambda$ decreases, the attribute prominence decreases) \cite{shen2020interpreting,denton2019image}. For our experiments, we use StyleGAN boundaries trained on the FFHQ dataset for gender and age. We also train our own boundary for ethnicity by grouping images into categories of majority and minority groups. Lastly, a quality boundary is used and applied after each attribute modification to remove image artifacts; we note that this boundary, provided by InterFaceGAN, was trained on the CelebA-HQ library.

\section{Ethical Considerations}\label{S-ethics}
\subsection{Intended Use}

The framework outlined in this paper is intended for analyzing visual biases in video-based predictor models. Our framework uses the OCEAN model, which is commonly applied in hiring and marketing contexts, and has been extensively tested and widely adopted in deep learning applications \cite{mehta2020recent}. However, we caution against the use of other personality models, such as Myers-Briggs, which, although used in some predictive systems, have been criticized for lacking empirical validity \cite{pittenger1993utility}. We also caution against applying this framework to other affective computing tasks, such as facial emotion inference, which has been shown to be unreliable due to cultural variation in emotional expression and the influence of social context and emotional masking \cite{barrett2019emotional}. Applying fairness auditing methods in such settings risks reinforcing invalid assumptions, as the underlying data and task definitions may already encode societal bias. Additionally, while our counterfactual modification framework can reveal and help quantify model biases, it is not intended as a prescriptive tool for real-time hiring decisions. We also emphasize that the protected attribute classifier used in this study serves only to evaluate the success of counterfactual generation. It is not used for any downstream prediction, selection, or inference task, and should not be repurposed for such applications. Fairness interventions must be interpreted in context and combined with broader policy and oversight mechanisms.

As noted by Denton et al. \cite{denton2019image}, bias measurement methods are only one part of building fair and inclusive AI systems. They are necessary but not sufficient. Broader systemic efforts—including transparent practices, participatory design, and regulatory oversight—are essential to meaningfully address the risks posed by facial analysis and automated decision-making systems. Moreover, fairness evaluations based on single-attribute comparisons may overlook intersectional disparities—where overlapping identities result in compounded disadvantages. We suggest this as a critical area for future research. While counterfactual fairness is a powerful auditing tool, we acknowledge its limitations when applied to immutable or socially sensitive attributes like ethnicity or gender, which cannot be meaningfully ``changed'' in reality.

\subsection{Dataset Limitations}
As described in Sec. \ref{S-dataset}, we use the CLFI dataset \cite{kaya2017multi}. Although this dataset is applicable to our affective computing research and has been used extensively in past work \cite{mehta2020recent}, we outline its limitations and why models developed from this paper are not to be used for real-time OCEAN trait prediction, but only for proof-of-concept. For this reason, we do not provide any pretrained models with our code.

First, this dataset contains labels of first impressions of OCEAN trait scores, created from pairwise annotations by MTurk workers, rather than from a personality assessment. This means these trait values do not reflect individuals' true personality traits or their performance in a job interview. Moreover, the videos are not real job interviews but rather 15-second clips where individuals are speaking to a camera, possibly about any subject. Second, we note limitations in the protected attribute labels of this dataset. There are three attributes given by this dataset: gender, ethnicity, and age. All attributes are perceived traits, labeled by MTurk workers; therefore, these do not necessarily reflect the true attributes of individuals but rather serve as a proof-of-concept to evaluate our framework.

Next, we note that this dataset is limited in its range of protected groups for each attribute; for instance, gender is labeled as binary, and ethnicity is labeled according to three groups (Caucasian, Asian, and Black). If used in real-world applications, this can reinforce rigid societal norms and harm individuals who exist outside of these groups. Lastly, the age attribute has only been labeled for individuals in the Caucasian ethnicity group; therefore, when using this label in any analysis, we are limited to a smaller amount of data and cannot observe intersections with other protected groups.

\section{Results and Discussion}\label{S-results}
In this section, we present the results of our three experiments. Due to ethical considerations outlined in Section~\ref{S-ethics}, we do not release any pretrained models. Our experiments evaluated multiple model variants to assess predictive accuracy, measure potential bias, and analyze the impact of counterfactual attribute manipulation on model fairness.

To assess the effect of different modalities and input types, we evaluated the following variants of our multi-task deep neural network (MTDNN) model:

\begin{itemize}

    \item \textbf{MTDNN}: The full model using all modalities, including visual, audio, and text transcript features.

    \item \textbf{MTDNN-Visual}: Model using only visual features extracted from video frames.

    \item \textbf{MTDNN-Audio}: Model using only audio features from the video’s soundtrack.

    \item \textbf{MTDNN-Text}: Model using only text-based features derived from video transcripts.

\end{itemize}

We also considered reduced-frame and GAN-based variations of these models:

\begin{itemize}

    \item \textit{Reduced-frame models}:
    \begin{itemize}
        \item \textbf{MTDNN-10F}: Full model using only the first 10 frames from each video.
        \item \textbf{MTDNN-Visual-10F}: Visual-only model using the first 10 frames.
    \end{itemize}

    \item \textit{GAN-inverted models} (using visual features derived from inversion to GAN latent space, without attribute manipulation):
    \begin{itemize}
        \item \textbf{MTDNN-Inv}: Full model using GAN-inverted visual features. These models serve as a reference to isolate the effects of GAN inversion (e.g., visual artifacts) from those of protected attribute modification.
        \item \textbf{MTDNN-Visual-Inv}: Visual-only version of the above.
    \end{itemize}

    \item \textit{Counterfactually modified models} (based on latent space editing to simulate different protected attributes):
    \begin{itemize}
        \item \textbf{MTDNN-Inv-Mod}: Full model using counterfactually modified video frames.
        \item \textbf{MTDNN-Visual-Inv-Mod}: Visual-only model using counterfactually modified video frames.
    \end{itemize}

\end{itemize}

In addition, for Experiment~3, we introduced a separate multi-task classification model:

\begin{itemize}
    \item \textbf{P-Attribute}: A model trained to predict perceived protected attributes (gender, ethnicity, and age) from video frames. This model was used to evaluate the effectiveness of our counterfactual video generation process.
\end{itemize}

This comprehensive set of models enabled us to analyze predictive accuracy, identify biases across demographic groups, and assess how counterfactual modifications influence fairness metrics.

\begin{figure*}[t!]
\centering
\includegraphics[width=1\textwidth,trim={0cm 0cm 0cm 0cm},clip]{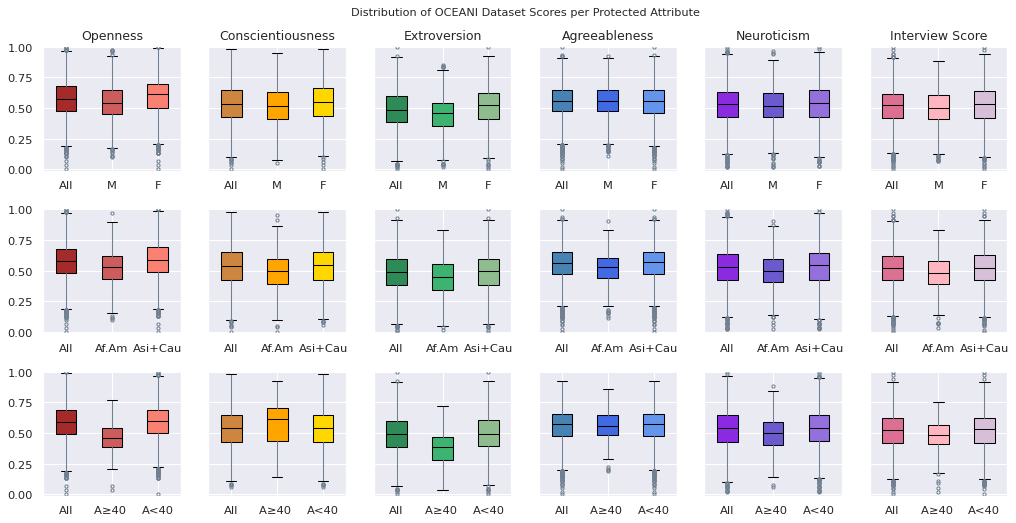}
\caption{Distribution of OCEANI dimension scores in the original dataset ground-truth labels, grouped by gender, ethnicity, and age. Each plot shows the distribution for all individuals, the protected group (``M'' for male, ``Af.Am.'' for African-American individuals, and ``$\text{A} \ge 40$'' for individuals aged 40 and above), and the unprotected group (``F'' for female, ``Asi.+Cau.'' for Asian and Caucasian individuals, and ``$\text{A} < 40$'' for individuals under 40).}
\label{fig:distribution_examples}
\end{figure*}

\begin{table*}[t]
  \centering
  \caption{Prediction performance of each model variant for individual OCEANI dimensions (left) and average performance across demographic groups (right), measured as (1 – MAE). Higher values indicate better performance.}
  \renewcommand{\arraystretch}{1.2}
  \begin{tabular}{lcccccc|ccccccc}
    \toprule
    \multirow{2}{*}{\textbf{Model}} &
    \multicolumn{6}{c|}{\textbf{OCEANI Dimension Prediction Performance}} &
    \multicolumn{7}{c}{\textbf{Demographic Group Prediction Performance}} \\[0.5ex]
    \cline{2-14}
    & \textbf{O} & \textbf{C} & \textbf{E} & \textbf{A} & \textbf{N} & \textbf{I} &
       \textbf{F}&\textbf{M}
      & \textbf{Asian} & \textbf{Af.Am.} & \textbf{Cau.} &  $\mathbf{A < 40}$ &  $\mathbf{A \ge 40}$\\
    \midrule
\textbf{MTDNN} & 0.911 & 0.911 & 0.916 & 0.911 & 0.910 & 0.913 & 0.909 &0.912 &0.912 &0.908 &0.911 &0.911 &0.911\\ 
\textbf{MTDNN-Visual} & 0.910 & 0.907 & 0.914 & 0.909 & 0.908 & 0.911 & 0.908 &0.910 &0.916 &0.905 &0.909 &0.909 &0.911\\ 
\textbf{MTDNN-Audio} & 0.897 & 0.890 & 0.893 & 0.900 & 0.894 & 0.895 & 0.893 &0.897 &0.905 &0.891 &0.895 &0.895 &0.887\\ 
\textbf{MTDNN-Text} & 0.883 & 0.874 & 0.877 & 0.893 & 0.877 & 0.880 & 0.877 &0.886 &0.890 &0.885 &0.879 &0.880 &0.877\\ 
\textbf{MTDNN-10F} & 0.903 & 0.902 & 0.902 & 0.902 & 0.900 & 0.903 & 0.901 &0.902 &0.907 &0.902 &0.901 &0.901 &0.902\\ 
\textbf{MTDNN-Visual-10F} & 0.893 & 0.889 & 0.893 & 0.896 & 0.886 & 0.890 & 0.892 &0.891 &0.900 &0.891 &0.892 &0.891 &0.892\\ 
\textbf{MTDNN-Inv} & 0.901 & 0.898 & 0.899 & 0.900 & 0.895 & 0.898 & 0.898 &0.899 &0.905 &0.897 &0.899 &0.899 &0.893\\ 
\textbf{MTDNN-Visual-Inv} & 0.892 & 0.885 & 0.891 & 0.892 & 0.882 & 0.886 & 0.888 &0.888 &0.897 &0.886 &0.888 &0.888 &0.890\\ 
  \bottomrule
  \end{tabular}
  \label{tbl:accuracy}
\end{table*}

\begin{table*}[t]
  \centering
  \caption{Mean and standard deviation of job interview score predictions for each model, expressed as $\mathcal{N}(\mu,\,\sigma)$ and grouped by demographic category. Demographic group abbreviations are consistent with those in Table~\ref{tbl:accuracy}.}
  \renewcommand{\arraystretch}{1.2}
  \begin{tabular}{lccccccc}
    \toprule
    \multirow{2}{*}{\textbf{Model}} &
    \multicolumn{7}{c}{\textbf{Job Interview Score Prediction Distribution $\mathcal{N}(\mu,\,\sigma)$ by Demographic Group}} \\[0.5ex]
    \cline{2-8}
    & \textbf{Female} & \textbf{Male} & \textbf{Asian} & \textbf{African American} & \textbf{Caucasian} &  \textbf{Age $ < 40$} &  \textbf{Age $ \ge 40$}\\
    \midrule
\textbf{Dataset Ground Truth}& 0.522, 0.152 &0.502, 0.143 &0.516, 0.136 &0.478, 0.139 &0.519, 0.150 &0.521, 0.150 &0.486, 0.143\\ 
\textbf{MTDNN}& 0.524, 0.112 &0.500, 0.108 &0.517, 0.100 &0.472, 0.102 &0.520, 0.112 &0.521, 0.112 &0.494, 0.111\\ 
\textbf{MTDNN-Visual}& 0.527, 0.106 &0.504, 0.104 &0.516, 0.099 &0.476, 0.101 &0.524, 0.105 &0.524, 0.105 &0.505, 0.099\\ 
\textbf{MTDNN-Audio}& 0.524, 0.090 &0.499, 0.088 &0.506, 0.088 &0.504, 0.085 &0.516, 0.090 &0.517, 0.091 &0.494, 0.085\\ 
\textbf{MTDNN-Text}& 0.507, 0.005 &0.507, 0.005 &0.507, 0.005 &0.507, 0.005 &0.507, 0.005 &0.507, 0.005 &0.507, 0.005\\ 
\textbf{MTDNN-10F}& 0.528, 0.109 &0.502, 0.100 &0.511, 0.104 &0.477, 0.098 &0.523, 0.106 &0.525, 0.106 &0.486, 0.097\\ 
\textbf{MTDNN-Visual-10F}& 0.527, 0.097 &0.499, 0.089 &0.512, 0.093 &0.473, 0.085 &0.522, 0.095 &0.524, 0.095 &0.491, 0.084\\ 
\textbf{MTDNN-Inv}& 0.526, 0.109 &0.499, 0.101 &0.522, 0.099 &0.478, 0.096 &0.520, 0.107 &0.521, 0.107 &0.484, 0.103\\ 
\textbf{MTDNN-Visual-Inv}& 0.526, 0.097 &0.500, 0.089 &0.521, 0.090 &0.480, 0.086 &0.520, 0.096 &0.522, 0.096 &0.480, 0.083\\ 
  \bottomrule
  \end{tabular}
  \label{tbl:distributions}
\end{table*}

\begin{table}[t]
  \centering
  \caption{Bias values for each model based on the independence criterion, measured for the interview score across gender, ethnicity, and age. Lower values indicate less dependence between model predictions and the protected attribute.}
  \renewcommand{\arraystretch}{1.2}
  \begin{tabular}{lccc}
    \toprule
    \multirow{2}{*}{\textbf{Model}} &
    \multicolumn{3}{c}{\textbf{Protected Group}} \\[0.5ex]
    \cline{2-4}
    & \textbf{Gender} & \textbf{Ethnicity} & \textbf{Age} \\
    \midrule
\textbf{Dataset}& 0.000 &0.007 &0.005 \\
\textbf{MTDNN}& 0.007 &0.011 &0.001 \\
\textbf{MTDNN-Visual}& 0.009 &0.010 &0.000 \\
\textbf{MTDNN-10F}& 0.004 &0.012 &0.004 \\
\textbf{MTDNN-Visual-10F}& 0.006 &0.016 &0.005 \\
\textbf{MTDNN-Inv}& 0.009 &0.004 &0.005 \\
\textbf{MTDNN-Visual-Inv}& 0.008 &0.010 &0.002 \\
\textbf{MTDNN-Inv-Mod}& 0.012 &0.012 &0.016 \\
\textbf{MTDNN-Visual-Inv-Mod}& 0.017 &0.008 &0.008 \\
  \bottomrule
  \end{tabular}
  \label{tbl:biases}
\end{table}

\begin{table}[t]
  \centering
  \caption{Disparate impact scores for each model across gender, ethnicity, and age, based on interview score–based candidate selection using a top-$N$ threshold of $N = 200$. A value of 1.0 indicates equal selection rates between protected and unprotected groups.}
  \renewcommand{\arraystretch}{1.2}
  \begin{tabular}{lccc}
    \toprule
    \multirow{2}{*}{\textbf{Model}} &
    \multicolumn{3}{c}{\textbf{Protected Group}} \\[0.5ex]
    \cline{2-4}
    & \textbf{Gender} & \textbf{Ethnicity} & \textbf{Age} \\
    \midrule
\textbf{Dataset}& 0.582 & 0.467 & 0.000 \\
\textbf{MTDNN}& 0.377 & 0.106 & 1.115 \\
\textbf{MTDNN-Visual}& 0.280 & 0.289 & 0.483 \\
\textbf{MTDNN-10F}& 0.260 & 0.365 & 0.239 \\
\textbf{MTDNN-Visual-10F}& 0.221 & 0.251 & 0.360 \\
\textbf{MTDNN-Inv}& 0.435 & 0.404 & 1.115 \\
\textbf{MTDNN-Visual-Inv}& 0.290 & 0.289 & 0.000 \\
\textbf{MTDNN-Inv-Mod}& 1.803 & 0.443 & 0.360 \\
\textbf{MTDNN-Visual-Inv-Mod}& 3.306 & 1.082 & 0.607 \\
  \bottomrule
  \end{tabular}
  \label{tbl:disparateimpact}
\end{table}

\begin{table}[t]
  \centering
  \caption{Disparate impact scores for each model across gender, ethnicity, and age, based on candidate selection using an interview score threshold of $\tau = 0.75$. A disparate impact score of 1.0 indicates equal selection rates between protected and unprotected groups, corresponding to statistical parity---a common form of group fairness.}
  \renewcommand{\arraystretch}{1.2}
  \begin{tabular}{lccc}
    \toprule
    \multirow{2}{*}{\textbf{Model}} &
    \multicolumn{3}{c}{\textbf{Protected Group}} \\[0.5ex]
    \cline{2-4}
    & \textbf{Gender} & \textbf{Ethnicity} & \textbf{Age} \\
    \midrule
\textbf{Dataset} & 0.586 & 0.464 & 0.091  \\
\textbf{MTDNN} & 0.260 & 0.122 & 1.820  \\
\textbf{MTDNN-Visual} & 0.274 & 0.612 & 0.000  \\
\textbf{MTDNN-10F} & 0.222 & 0.243 & 0.000  \\
\textbf{MTDNN-Visual-10F} & 0.202 & 0.990 & 0.000  \\
\textbf{MTDNN-Inv} & 0.199 & 0.220 & 0.394  \\
\textbf{MTDNN-Visual-Inv} & 0.182 & 1.312 & 0.000  \\
\textbf{MTDNN-Inv-Mod}& 2.088 & 0.770 & 0.394 \\
\textbf{MTDNN-Visual-Inv-Mod}& 4.577 & 3.186 & 1.314  \\
  \bottomrule
  \end{tabular}
\label{tbl:disparateimpactthreshold}
\end{table}

\subsection{Experiment 1: Personality Prediction Model Results}
In Experiment~1, we present the accuracy results of the MTDNN model and its variants, analyzing performance across different input modalities (e.g., visual, audio), the distribution of OCEANI scores across protected groups, and the presence of initial bias in these models. These results provide insight into potential biases in the dataset and model predictions, as well as general trends in OCEANI dimension scores.

First, we examine the prediction performance per OCEANI dimension and the average performance across demographic groups for the baseline MTDNN model, as shown in Table~\ref{tbl:accuracy} on the left and right sides, respectively. These values are computed as (1 – mean absolute error), where higher values indicate better performance. From these results, we make the following observations: (a) The MTDNN model achieves the highest prediction performance across all OCEANI dimensions and demographic groups compared to the other model variants. (b) Among models that use a single modality, performance ranks in the order: visual, audio, and text, with the visual-only model achieving the highest prediction performance and the text-only model the lowest, though the latter still achieves a (1 – MAE) score above 87\% across all dimensions. (c) There is a slight decrease in prediction performance when using only the first 10 frames of video, but performance remains high. (d) The GAN-inverted models (MTDNN-Inv and MTDNN-Visual-Inv) perform well, though slightly below their original counterparts. (e) Across all demographic groups, average prediction performance is comparable between majority and minority groups, with no large disparities observed.

To further investigate disparities in model outputs, we analyzed the distribution of interview scores across demographic groups, as shown in Table~\ref{tbl:distributions} and Figure~\ref{fig:distribution_examples}.

We also present the distribution of job interview scores across protected groups in Table~\ref{tbl:distributions}, and box plots of all OCEANI dimensions in Figure~\ref{fig:distribution_examples}. We focus on the interview score because it aggregates information from all OCEAN traits and represents a practical output of affective computing models in real-world hiring scenarios. From these results, we observe that mean and median interview scores tend to differ between majority and minority groups. In particular, the protected groups—male, older individuals ($\text{A} \ge 40$), and African-American individuals—consistently have lower median scores than the overall dataset median, as shown in Figure~\ref{fig:distribution_examples}. Table~\ref{tbl:distributions} similarly shows small but consistent disparities in mean interview scores. For example, individuals under the age of 40 have a slightly higher average interview score than those aged 40 or older. Likewise, Asian and Caucasian individuals have similar mean scores, both slightly higher than those of the African-American group. These differences in interview score distributions suggest potential bias in the ground-truth labels themselves, which could arise from several factors, including qualitative differences in representation, rater bias, societal stereotypes, or socioeconomic disparities that affect perceived performance.

\begin{figure}[t]
\centering
\includegraphics[width=.47\columnwidth,trim={0cm 0cm 0cm 0cm},clip]{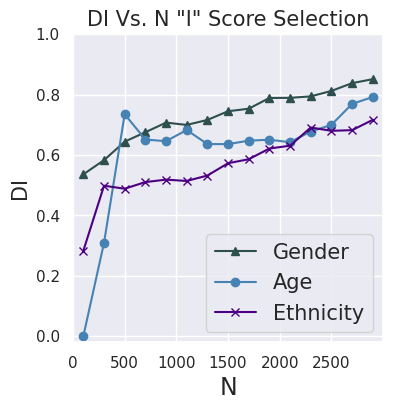}\quad
\includegraphics[width=.47\columnwidth,trim={0cm 0cm 0cm 0cm},clip]{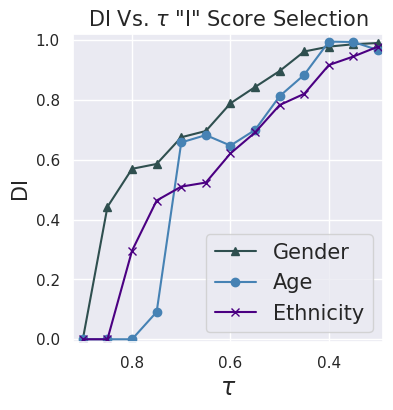}
\caption{Disparate impact (DI) scores for the MTDNN model as a function of candidate selection criteria, using interview score as the selection variable. The left plot shows disparate impact versus top-$N$ selection; the right plot shows disparate impact versus $\tau$ threshold selection.}
\label{fig:DIexamples}
\end{figure}

Finally, we analyze model bias using established fairness metrics. Table~\ref{tbl:biases} reports mutual information scores to assess the statistical independence between interview score predictions and protected attributes. Tables~\ref{tbl:disparateimpact} and~\ref{tbl:disparateimpactthreshold} present disparate impact scores based on interview score–driven candidate selection, using two common approaches: top-$N$ selection and score thresholding. To identify thresholds that reveal the most pronounced bias, we experimented with several values and selected $N = 200$ and $\tau = 0.75$ for our analyses. As shown in Figure~\ref{fig:DIexamples}, these thresholds result in disparate impact values that fall significantly below 1.0, indicating strong selection bias in the original dataset labels. As selection becomes more inclusive—either by increasing $N$ or lowering $\tau$—the disparate impact gradually approaches 1.0, reflecting more equitable outcomes. However, since real-world hiring decisions typically involve selecting a limited number of candidates, we adopt $N = 200$ and $\tau = 0.75$ as standard parameters for comparing model bias across modalities and architectures.

In all tables presenting bias results, we include scores for both the full MTDNN model and its visual-only counterpart, MTDNN-Visual, which uses only facial features and excludes scene information. This comparison allows us to evaluate how visual inputs contribute to model bias. As shown in Table~\ref{tbl:biases}, moving from the dataset ground truth to the MTDNN model generally leads to increased dependency between protected attributes and model predictions, as reflected by higher mutual information scores. Although these scores remain relatively small across all groups, they indicate a measurable shift. Similarly, as shown in Table~\ref{tbl:disparateimpactthreshold}, disparate impact scores often decline when moving from the ground truth to the MTDNN and MTDNN-Visual models—particularly for gender—indicating greater bias in certain cases when using visual features alone.

\begin{figure*}[t]
\centering
\includegraphics[width=\textwidth,trim={0cm 0cm 0cm 0cm},clip]{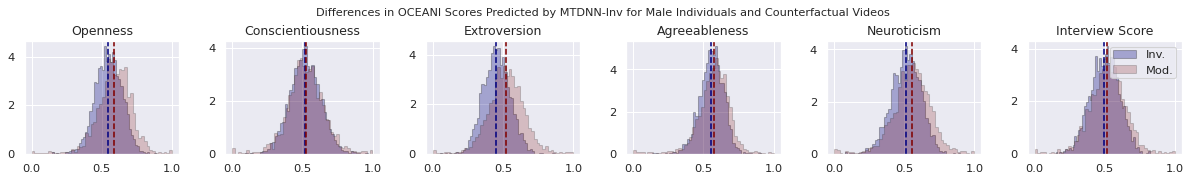}\quad
\includegraphics[width=\textwidth,trim={0cm 0cm 0cm 0cm},clip]{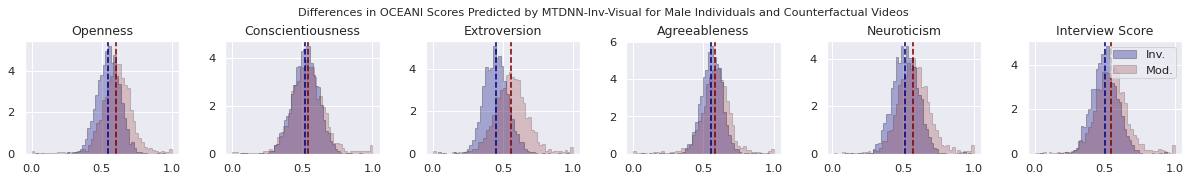}
\caption{Distributions of predicted OCEANI dimension scores for individuals and their counterfactual counterparts. The first row shows predictions for male individuals using the full model and the changes in scores when their counterfactual videos are input. The second row shows predictions using the visual-only model (MTDNN-Visual-Inv) for the same individuals and their counterfactuals.}
\label{fig:distributions_plot}
\end{figure*}

\begin{figure}[t]
\centering
\includegraphics[width=.3\columnwidth]{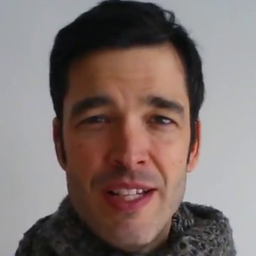}\quad
\includegraphics[width=.3\columnwidth]{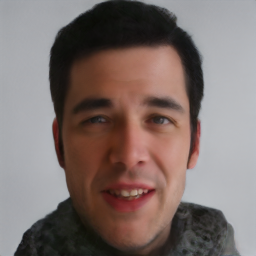}\quad
\includegraphics[width=.3\columnwidth]{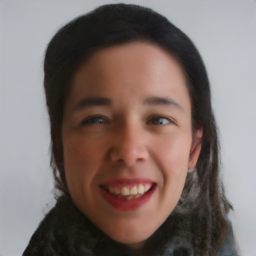}

\medskip
\includegraphics[width=.3\columnwidth]{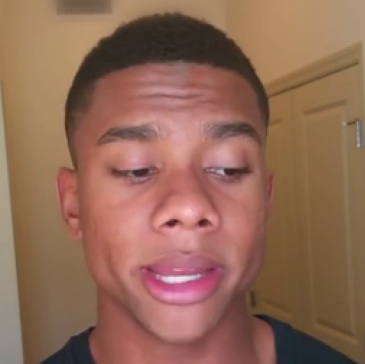}\quad
\includegraphics[width=.3\columnwidth]{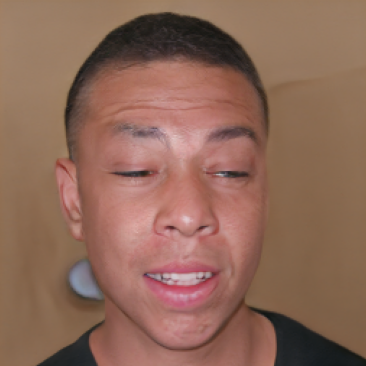}\quad
\includegraphics[width=.3\columnwidth]{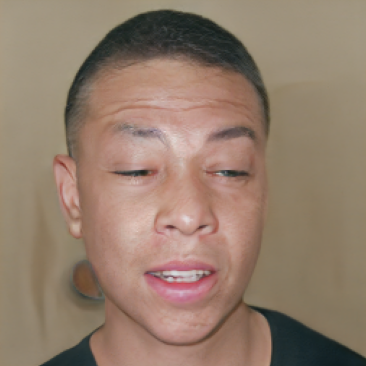}

\medskip
\includegraphics[width=.3\columnwidth]{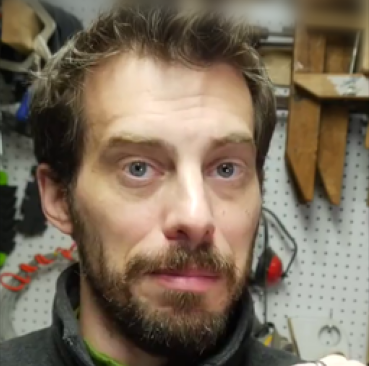}\quad
\includegraphics[width=.3\columnwidth]{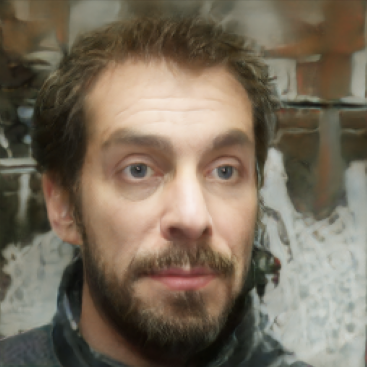}\quad
\includegraphics[width=.3\columnwidth]{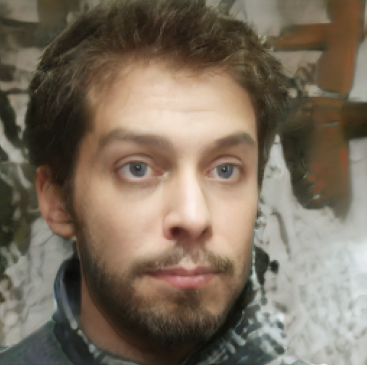}

\caption{Example video frames showing the original, GAN-inverted, and counterfactually modified images for three individuals. For each row, the first image is the original frame, the second is the GAN-inverted reconstruction, and the third is the counterfactual frame generated by manipulating a protected attribute via latent space boundary editing (first row: gender changed from male to female; second row: ethnicity changed to Caucasian; third row: age reduced for a male individual).}
\label{fig:example_photos}
\end{figure}

\subsection{Experiment 2: Counterfactual Analysis}

\begin{figure*}[t!]
\centering
\includegraphics[width=1\textwidth,trim={0cm 0cm 0cm 0cm},clip]{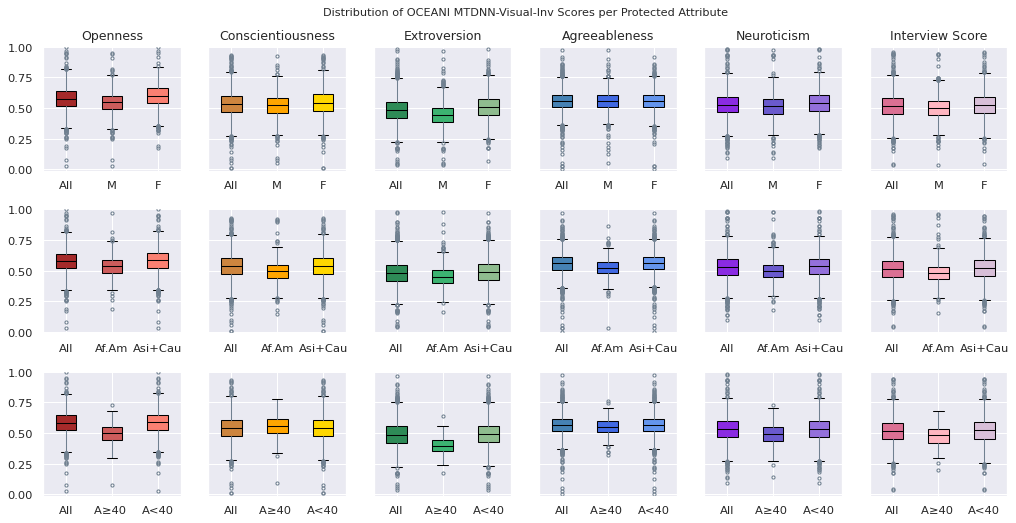}
\caption{Distributions of OCEANI dimension scores predicted by our MTDNN-Visual-Inv model for individuals across different genders, ethnicities, and age groups. Each plot shows the distribution for all individuals, the protected group (``M'' for male, ``Af.Am.'' for African-American individuals, and ``$\text{A} \ge 40$'' for individuals aged 40 and above), and the unprotected group (``F'' for female, ``Asi.+Cau.'' for Asian and Caucasian individuals, and ``$\text{A} < 40$'' for individuals under 40).}
\label{fig:distribution_inv}
\end{figure*}

In this experiment, we evaluate our reduced-frame (MTDNN-10F) and GAN-inverted (MTDNN-Inv) models. 
Results for (1 – MAE) prediction scores, score distributions by protected group, and bias metrics are reported in the same tables as Experiment~1.
We observe a slight decrease in interview score prediction performance compared to the original MTDNN and MTDNN-Visual models: the original MTDNN achieved 91.3\% prediction performance, while MTDNN-10F achieved 90.3\%. The MTDNN-Visual-10F model, which relies solely on visual frames, produced the lowest prediction performance at 89.0\%, likely due to the limited feature set. The GAN-inverted models, MTDNN-Inv and MTDNN-Visual-Inv, achieved similar prediction performance across traits and demographic groups compared to their original counterparts. This indicates that the inversion process introduces minimal distortion and does not significantly affect model predictions, making it a suitable baseline for counterfactual analysis. The predicted score distributions by protected attribute group for MTDNN-Visual-Inv are shown in Figure~\ref{fig:distribution_inv}. When compared to the original dataset distributions (Figure~\ref{fig:distribution_examples}), we observe that disparities between protected and unprotected groups persist in the model’s predictions—not only in the interview score, but also across OCEAN dimensions. For instance, protected groups consistently received lower average scores in traits such as extraversion and openness, which are positively correlated with the interview score and may negatively affect downstream selection outcomes.

\begin{figure*}[t!]
\centering
\includegraphics[width=1\textwidth,trim={0cm 0cm 0cm 0cm},clip]{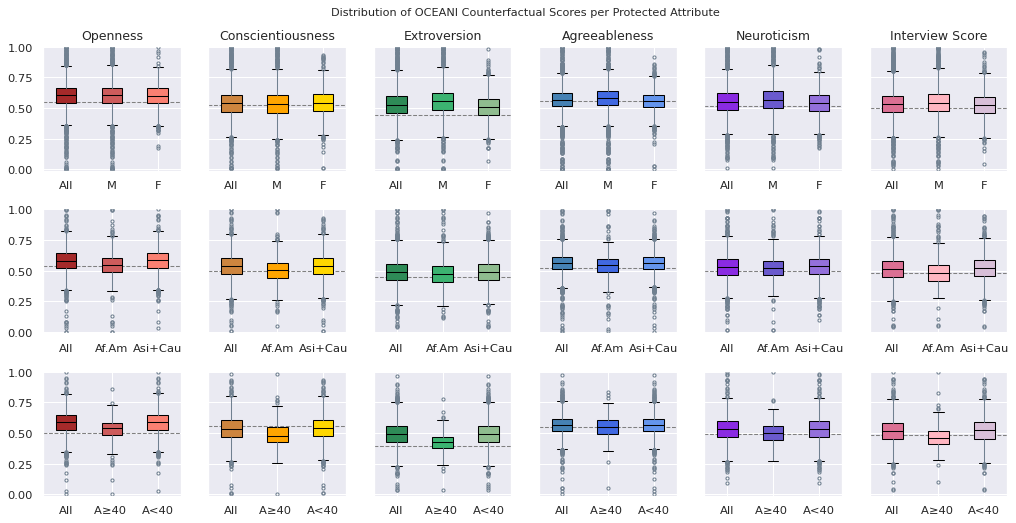}
\caption{Counterfactual predictions of OCEANI dimension scores for individuals in protected groups, following latent space attribute modification using the MTDNN-Visual-Inv model. Each plot shows the predicted score distribution for all individuals, the protected group (``M'' for male, ``Af.Am.'' for African-American individuals, and ``$\text{A} \ge 40$'' for individuals aged 40 and above), and the unprotected group (``F'' for female, ``Asi.+Cau.'' for Asian and Caucasian individuals, and ``$\text{A} < 40$'' for individuals under 40). For protected group members, the plotted scores are derived from their counterfactual videos in which the protected attribute was modified to match the corresponding unprotected group. The horizontal grey line in each plot is the median value from the corresponding protected group distribution in Figure~\ref{fig:distribution_inv}.}
\label{fig:distribution_counterfactual}
\end{figure*}

Next, we modify videos of individuals from protected groups using our counterfactual generation approach (Section~\ref{S-GAN}) and input the resulting counterfactual videos into the previously described models to examine how predictions change when protected attributes are altered. Examples of original, GAN-inverted, and counterfactually modified video frames are shown in Figure~\ref{fig:example_photos}. Figure~\ref{fig:distribution_counterfactual} presents the distribution of OCEANI dimension scores predicted for these modified videos. Compared to the original predictions shown in Figure~\ref{fig:distribution_inv}, the median scores of protected groups shift closer to those of their unprotected counterparts. This suggests that the counterfactual modifications effectively reduce prediction disparities. In some cases, such as gender, the modified group’s scores even slightly exceed those of the original unprotected group. For example, male individuals originally exhibited lower median extraversion scores compared to females, and are considered the protected group here for fairness assessment purposes due to their lower representation in the training data; after modifying their videos to reflect the female attribute, their predicted extraversion scores increased to match—and in some cases slightly surpass—the female group’s median. A similar trend is observed for ethnicity: African-American individuals originally received lower scores in traits such as openness and extraversion, but after being counterfactually modified to appear as unprotected group (Asian and Caucasian) members, their predicted scores increased accordingly, though slightly. Similarly, individuals over the age of 40 saw increases in openness and extraversion. These results suggest that the model’s original predictions were influenced by protected attributes, and that counterfactual modifications effectively reveal such underlying biases. We also observe potential negative effects when performing counterfactual modifications—for example, a decrease in conscientiousness for individuals over the age of 40, and an increase in neuroticism for male and African-American individuals. These changes may negatively impact the job interview process, as the counterfactual group shifts cause individuals’ traits to align more closely with those of the opposite group, which originally exhibited higher or lower scores in those traits.

We next evaluate the effect of counterfactual input modification on model bias by comparing predictions from the original inverted models (MTDNN-Inv and MTDNN-Visual-Inv) to those from their counterfactually modified versions (MTDNN-Inv-Mod and MTDNN-Visual-Inv-Mod). The impact of these modifications is summarized in Table~\ref{tbl:di_summary}, which reports disparate impact values before and after counterfactual modification for each protected attribute, under both top-$N$ and threshold-based selection strategies. As shown, counterfactual modification leads to substantial improvements in disparate impact across gender, ethnicity, and age. For example, under top-$N$ selection ($N = 200$), the DI value for gender increases from 0.290 to 3.306, for ethnicity from 0.289 to 1.082, and for age from 0.000 to 0.607. Under threshold-based selection ($\tau = 0.75$), gender DI increases from 0.182 to 4.577, ethnicity from 1.312 to 3.186, and age from 0.000 to 1.314. These improvements indicate more equitable candidate selection rates after counterfactual modification. The effects are also visualized in Figure~\ref{fig:distributions_plot}, which specifically shows predicted OCEANI scores for male individuals before and after modification. For these individuals, counterfactual edits led to increases in interview scores and related personality traits—such as openness, conscientiousness, and extraversion—all of which are positively associated with interview outcomes. Similar score improvements were observed for African-American individuals and for individuals over the age of 40, contributing to the overall fairness gains observed in our evaluation. This suggests that counterfactual modification helped boost scores for several individuals in the minority group (specifically African-American individuals), including some outliers who received notably high interview scores, as seen in Figure~\ref{fig:distributions_plot}. This behavior contributes to the high post-modification disparate impact values in Table~\ref{tbl:di_summary}, as score shifts for a subset of protected individuals—particularly those near the selection threshold—led to disproportionately higher selection rates. While some disparate impact values exceed 1.0, potentially suggesting over-correction, we emphasize that our approach is intended for fairness auditing rather than deployment. In this context, DI values greater than 1.0 are not inherently problematic; they indicate that counterfactual transformation shifted some individuals’ scores beyond the original group distribution, which is expected in a sensitivity analysis framework. These results highlight the model's sensitivity to protected attributes and demonstrate the effectiveness of counterfactual modification in reducing prior disparities.

\begin{table}[t]
  \centering
  \caption{Disparate impact (DI) before and after counterfactual modification, evaluated using two candidate selection strategies: top-$N$ selection ($N = 200$) and threshold-based selection ($\tau = 0.75$). DI values are drawn from Tables~\ref{tbl:disparateimpact} and~\ref{tbl:disparateimpactthreshold}, comparing the MTDNN-Visual-Inv model (before) and MTDNN-Visual-Inv-Mod model (after). DI values closer to 1.0 indicate more equitable selection outcomes. Values significantly above or below 1.0 signal imbalance.}
  \label{tbl:di_summary}
  \renewcommand{\arraystretch}{1.2}
  \setlength{\tabcolsep}{5.3pt}
  \begin{tabular}{lcccc}
    \toprule
    \multirow{2}{*}{\textbf{Group}} &
    \multicolumn{2}{c}{\textbf{Top-$N$ ($N = 200$)}} &
    \multicolumn{2}{c}{\textbf{Threshold ($\tau = 0.75$)}} \\
    \cmidrule(lr){2-3} \cmidrule(lr){4-5}
    & \textbf{Before} & \textbf{After} & \textbf{Before} & \textbf{After} \\
    \midrule
    Gender    & 0.290 & 3.306 & 0.182 & 4.577 \\
    Ethnicity & 0.289 & 1.082 & 1.312 & 3.186 \\
    Age       & 0.000 & 0.607 & 0.000 & 1.314 \\
    \bottomrule
  \end{tabular}
\end{table}

\subsection{Experiment 3: Evaluation of Counterfactual Generation} 

In Experiment~3, we evaluate the effectiveness of our counterfactual video generation framework by assessing whether attribute modifications successfully alter the perceived protected attribute in a given video. To do so, we use a separately trained model (Section~\ref{S-attribute-model}), referred to as the \textbf{P-Attribute} model, to classify perceived gender, ethnicity, and age from video frames. We then input counterfactual videos—generated by modifying a protected attribute—into this model. If the counterfactual manipulation is effective, the model should misclassify the protected attribute, indicating that the counterfactual now resembles the unprotected group.

Results are shown in Table~\ref{tbl:exp3}. The P-Attribute model achieved high classification accuracy on original video frames: 85.5\% for gender, 84.5\% for ethnicity, and 91.9\% for age. Additionally, it achieved relatively high per-class recall across protected groups. After counterfactual modification, we observe a notable decrease in recall for the targeted protected classes, indicating that the perceived attribute has shifted.

For example, when modifying videos of individuals aged $A \ge 40$ to appear younger, the recall for that class dropped from 0.85 to 0.17, with the model now predicting the individuals as belonging to the $A < 40$ group. Similarly, for the Male group, recall decreased from 0.85 to 0.23 after modifying the videos to reflect the female attribute. Smaller recall reductions were observed in other classes, such as the Female group (0.93 to 0.61) and the African-American group (0.92 to 0.68), indicating the attribute shift was still present, though less pronounced.

Overall, these results demonstrate that our counterfactual generation framework successfully modifies the protected attributes in a way that alters the model’s perception. This validates the realism and effectiveness of our latent space manipulation technique for generating counterfactuals.

\begin{table}[t]
  \centering
  \caption{Recall per class of the P-Attribute model on original video frames (``Orig.'') and counterfactual video frames (``Count.''). Counterfactual videos were generated by modifying individuals in protected groups to match the corresponding unprotected group. A drop in recall for a protected class after modification indicates successful attribute manipulation. Cells marked ``--'' correspond to classes that were not targeted for modification.}
  \label{tbl:exp3}
  \renewcommand{\arraystretch}{1.2}
  \setlength{\tabcolsep}{5.3pt}
  \begin{tabular}{lccccccc}
    \toprule
    \textbf{Input} & \multicolumn{2}{c}{\textbf{Gender}} & \multicolumn{3}{c}{\textbf{Ethnicity}} & \multicolumn{2}{c}{\textbf{Age}} \\
    \cmidrule(lr){2-3} \cmidrule(lr){4-6} \cmidrule(lr){7-8}
    & F & M & Af.Am. & Cau. & Asi. & $\text{A} < 40$ & $\text{A} \ge 40$ \\
    \midrule
    \textbf{Orig.}  & 0.93 & 0.85 & 0.92 & 0.87 & 0.83 & 0.96 & 0.85 \\
    \textbf{Count.} & 0.61 & 0.23 & 0.68 & --   & 0.57 & --   & 0.17 \\
    \bottomrule
  \end{tabular}
\end{table}

\section{Conclusion}\label{S-conclusion}
We presented a framework for measuring bias in affective computing models through counterfactual manipulation of protected attributes. Our approach leverages StyleGAN to generate facial image modifications along a single attribute boundary, enabling controlled edits of input videos. By comparing model predictions on original and counterfactual inputs, we quantified how outputs change when individuals shift between protected and unprotected groups. We applied fairness metrics, including statistical independence and disparate impact (as a proxy for demographic parity), to assess model bias before and after counterfactual modification. Additionally, we trained a protected attribute classifier to validate the effectiveness of our counterfactual generation process. This classifier exhibited significant drops in classification accuracy when evaluated on counterfactual videos, confirming that the GAN successfully altered perceived protected attributes. Importantly, because our method operates entirely on model inputs and outputs, it enables fairness assessments even when the underlying AI system is a black box—an increasingly common scenario in industry.

We identify several directions for future work. First, while our study focuses on visual attributes, the MTDNN model incorporates multimodal features. Extending counterfactual modification to audio (e.g., voice pitch, tone) and text (e.g., speech content) would enable a more comprehensive evaluation of multimodal bias. Second, instead of modifying individual frames, future work could explore the use of generative models designed for full video synthesis to improve temporal consistency and realism. Finally, we recommend applying this framework to more diverse and representative datasets. The CLFI dataset, while widely used, is limited to short clips and a small number of protected attribute categories. 
Future work should also examine intersectional bias, where overlapping identities (e.g., gender and ethnicity) may lead to compounded disadvantages that are not captured by single-attribute analysis. Expanding to real job interview data and underrepresented groups would provide deeper insights into fairness in affective computing.

Overall, our work introduces a generalizable and practical framework for fairness auditing in video-based AI systems. It can support researchers and practitioners working in affective computing, hiring, education, and other domains where fairness and accountability are critical. Our results show that counterfactual modification can substantially improve fairness outcomes, as demonstrated by increased disparate impact values across gender, ethnicity, and age. This underscores the utility of our framework for identifying and quantifying hidden biases in AI systems—especially in settings where AI systems are opaque and only input-output access is possible.

\IEEEpeerreviewmaketitle

\ifCLASSOPTIONcompsoc
  \section*{Acknowledgments}
\else
  \section*{Acknowledgment}
\fi
This material is based upon work supported by the National Science Foundation under Grant No. 1936857.

\ifCLASSOPTIONcaptionsoff
  \newpage
\fi

\bibliographystyle{IEEEtran}
\bibliography{refs}

\end{document}